# Graphene wrinkling induced by monodisperse nanoparticles: facile control and quantification

Jana Vejpravova[1]*, Barbara Pacakova[1], Jan Endres[2], Alice Mantlikova[1], Tim Verhagen[1], Vaclav Vales[3], Otakar Frank[3] and Martin Kalbac[3]**

[1]Institute of Physics AS CR, v.v.i., Department of Magnetic Nanosystems, Na Slovance 2, 18221 Prague 2, Czech Republic

[2]Charles Univeristy in Prague, Faculty of Mathematics and Physics, Department of Condensed Matter Physics, Ke Karlovu 5, 12116 Prague 2, Czech Republic

[3]JH Institute of Physical Chemistry AS CR,v.v.i., Dolejskova 3, 18200 Prague 8, Czech Republic

*vejpravo@fzu.cz, **martin.kalbac@jh-inst.cas.cz

*Abstract:* Controlled wrinkling of single-layer graphene (1-LG) at nanometer scale was achieved by introducing monodisperse nanoparticles (NPs), with size comparable to the strain coherence length, underneath the 1-LG. Typical fingerprint of the delaminated fraction is identified as substantial contribution to the principal Raman modes of the 1-LG (G and G'). Correlation analysis of the Raman shift of the G and G' modes clearly resolved the 1-LG in contact and delaminated from the substrate, respectively. Intensity of Raman features of the delaminated 1-LG increases linearly with the amount of the wrinkles, determined by advanced processing of atomic force microscopy data. Our study thus offers universal approach for both fine tuning and facile quantification of the graphene topography up to ~ 60% of delamination.

Control of graphene topography at nanoscale belongs to hot topics in condensed matter physics[1,2] and numerous areas of applied research, such as strain engineering targeting graphene-based electronics[3,4], sensing of mechanical fields and various kinds of molecules[5-9] or development of novel biomedical platforms[10,11]. Opening of a band gap and its fine control via external physical parameters mirrored in the spatial distribution of the strain and doping still remains a challenging task and one of the most discussed issues in graphene-related research[12-15]. In spite of the fact, that the elastic properties of the graphene are mainly related to the σ bonds (responsible for rigidity of the structure), the strain also strongly affects the π bonds enhancing the reactivity of the graphene as the π-orbitals become destabilized[16]. Therefore local control of the graphene topography opens door to enhance chemical reactivity of the graphene with spatial selectivity at nanometer scale[17].

The topography of graphene transferred on a substrate of interest is characteristic by network of topographic aberrations with a specific curvature, termed wrinkles. They form as a consequence of difference in thermal expansion of the graphene and substrate used for the growth, partial replication of the substrate topography[18] or they can be induced during the transfer process itself[19]. Their formation is also connected with induction of shear strain resulting in specific changes of the electronic structure[20]. Consequently, the density and mobility of the charge carriers in wrinkles show significant difference with respect to the ideally flat graphene sheet[2,21]. Even the simplest corrugation leading to inhomogeneous strains (periodic array of ripples) causes formation of non-trivial gauge fields acting on the charge carriers[1,2,22]. On the other hand, the corrugations in the substrate induce stresses, which can give rise to mechanical instabilities and the formation of wrinkles[22]. Therefore strategies to control the graphene topography may simply profit from natural occurrence of wrinkles by sophisticated tuning of the wrinkling process itself.

The wrinkles in graphene can be induced in graphene transferred on elastic corrugated substrates[23] or on substrates patterned with different nanoobjects[24-27]. The later approach using NPs seems to be rather promising; however the particles size must be comparable to the typical strain coherence length[28, 29]. Moreover, extremely narrow particle size distribution is desirable to control the wrinkling mechanism efficiently.

In our work we focused on control of wrinkling of 1-LG with the help of strictly monodisperse nanoparticles NPs, which size is comparable to the characteristic strain coherence length in graphene on $SiO_2$ substrate[28, 29]. The 1-LG@NPs nanostructures with different level of wrinkling were obtained by transfer of the 1-LG grown by chemical vapor deposition (CVD) on $SiO_2$/Si substrate decorated with different concentration of 9 nm NPs (schematically shown in Figure 1) and characterized by atomic force microscopy (AFM), scanning electron microscopy with high resolution (HR SEM) and Raman micro-spectroscopy. The AFM data were processed in order to determine the relative area of the delaminated 1-LG, attributed to the spine of the wrinkles. The Raman spectra of 1-LG@NPs revealed the essential characteristics related to the modification of lattice and electronic structure, reflected in the peak parameters of the principal Raman active modes of 1-LG[30-34]. The spectra were further subjected to correlation analysis, which enabled evaluation of the doping and strain in the delaminated and flat (strictly speaking fraction with roughness comparable to the substrate) parts of the 1-LG. Finally, we put in context the results of AFM and Raman data analysis and obtained a general linear dependence between the delaminated area of the 1-LG and Raman intensity of the related subbands. This general correlation serves as essential proof of robust control of the wrinkling in the 1-LG@NPs nanostructures.

Using the methodology presented in Figure 1, we succeeded in preparation of samples with $N_{NPs}$ ranging from 20 to ~ 500 NPs/µm². Typical AFM and HR SEM images of the 1-LG@NPs samples are shown in Figure 2; those of the NP-decorated substrates are included in Supplementary (Figure S3.1). The micrographs confirmed homogeneous distribution of the NPs on the substrates. Statistical analysis of the microscopic data provided the mean particle size ($d_{NP}$), distance ($d_{NP-NP}$) and surface density ($N_{NPs}$), which were in excellent agreement with the values predicted by the simulation for most of the samples; the results are summarized in Table S3.1. The root-mean square roughness, $\sigma_p$ and other important characteristic of the 1-LG@NPs samples, was derived from the AFM data (for details, see [35]). Finally, the relative area of the delaminated 1-LG, $A_w$ was obtained for each sample as projected surface area of the wrinkles resolved by AFM with the help of statistical grain analysis and triangulation procedure[35]. The $A_w$ is the important parameter used in generalized correlation of the AFM and Raman data as demonstrated further. We also observed that the $A_w$ roughly scales with $N_{NPs}$ and $1/d_{NP-NP}$ (Figure S3.2).

The key technique for analysis of the doping and strain of the 1-LG@NPs is the Raman spectroscopy. Typical Raman spectra within the G and G' regions are shown in Figure 2. The G mode shows a fine structure with rather clear contributions of two subbands, while the G' appears as asymmetric and rather broad band with complex inner structure. Considering the nature of our samples and appearance of the principal bands, the data were then analyzed as follows: the D and D' modes were sufficiently described using a single pseudo-Voigt function, while the G and G' modes were assumed as superposition of two pseudo-Voigt contributions. The two contributions recognized in the G and G' are then termed as $G_1$, $G_2$ and $G'_1$, $G'_2$. Typical decomposition of the G and G' band for all samples is shown in Figure 2, representative fits of the D band can be found in Supplementary (Figure S1.1). As the inner structure of G' mode of 1-LG on solid substrates is expected to be rather complex due to distribution of strain [14, 15, 30], in the first attempt to analyze the profile a single pseudo-Voigt function with large FWHM was used; the admixture of the Gaussian component is assumed as the measure of Raman shift distribution within the probed region (given by the laser spot size). Due to the obvious asymmetry of the G' peak, the simplified description was not sufficient (except for about half of the spectra in the most wrinkled sample - GNP3 which will be discussed later); the G' band was decomposed to $G'_1$, $G_2'$ in analogy with the G band, where the bimodal character is obvious and it is also expected to be of the similar origin.

Note that the G' of the GNP3 (with the highest level of wrinkling) can be reasonably fitted by using a single pseudo-Voigt function with large FWHM (~ 45 cm$^{-1}$), which reflects substantial roughness of the substrate[36]. However; keeping the same fitting model as in case of other samples, about 30 % of the spectra can be successfully analyzed by the two components $G'_1$ and $G'_2$.

Representative maps of the refined values of the integral Raman intensity and Raman shift of the $G_1$, $G_2$, $G'_1$ and D modes are shown in Figure 3 for the GNP3 sample, results for other samples are included in Supplementary, Figures S1.2 – S1.6. Both parameters show very good homogeneity within the probed region suggesting that the spatial variation of the strain and doping due to formation of wrinkles occurs at much lower scale, which is comparable to the dimension of wrinkles and below dimensions of the laser spot. Median values of the Raman intensities of the $G_1$, $G_2$, $G'_1$ and $G_2'$ modes were evaluated to be used for general correlation to the AFM data. Finally, median values of the other profile parameters corresponding to the particular bands were obtained and summarized in Table S1.2.

The doping ($n$) and strain ($\varepsilon$) related change of the Raman shift of the G and G' modes was assessed by applying the correlation analysis proposed by Lee et al[28]. Figure 4 represents the correlation diagram of the Raman shifts of the G' and G components for all samples. The two sets of correlation pairs ($G_1$, $G'_1$ and $G_2$, $G'_2$) for each sample are plotted with the same color. The diagram can be decomposed to series of iso-strain (doping) and iso-doping (strain) lines; those for neutral and unstrained graphene intersect at point $P_0$ (1582, 2637) cm$^{-1}$ and separate the correlation diagram to regions of allowed and physically irrelevant values. Each point in the correlation diagram can thus be described as a linear combination of the unit vectors corresponding to the strain, $\mathbf{e}_\varepsilon$ and doping, $\mathbf{e}_n$ with the origin of $P_0$.[37]

The Raman shift of G ($\omega_G$) and G' ($\omega_{G'}$) are highly sensitive to both $n$ and $\varepsilon$ with very different fractional variation, ($\Delta\omega_{G'}/\Delta\omega_G$) due to $n$ and $\varepsilon$. For low charge carrier concentrations, the shift of $\omega_{G'}$ ($\Delta\omega_{G'}$) as a function of $n$ is negligible, compared to the change in $\omega_G$ ($\Delta\omega_G$), and can be estimated within the iso-strain line. For biaxial strain, the slope of the iso-doping (strain) line was reported in the range 2.25 - 2.8[38-40]. In our samples, the 1-LG is transferred on substrate with artificial coarseness in scale of few nanometers. Since the strain coherence length can be also as small as a few nanometers for graphene on SiO$_2$ substrates[29], we obtain very different local strain contributions, which results in complex inner structure of the G' mode (and its subbands $G'_1$ and $G'_2$) enveloped by the experimental curve. Although deeper analysis of the G' is not possible with the available spatial resolution of Raman spectrometer, we can still assess the size of strain within iso-doping line (slope ~ 0.7) considering biaxial strain (slope of 2.45) and $\Delta\omega_{G'}$ about -144 cm$^{-1}$/1%[39].

The doping can be then estimated from the $\Delta\omega_G$ within the iso-strain line (slope ~ 2.45) using the formula proposed by Das et al [41]. In our samples, hole doping is expected as suggested by numerous transport experiments carried out on 1-LG transferred on SiO$_2$/Si substrate[42].

In the correlation diagram, the G;G' pairs group in two dominant regions labelled as $G_1$;$G'_1$ and $G_2$;$G'_2$. The data points of the $G_1$;$G'_1$ pairs of the GNP1, GNP2, GNP5 and GNP6 samples with $A_w$ up to ~ 20 % coincide in almost a single symmetric spot matching the line of biaxial strain of ~ -0.10% with doping about 1.3×10$^{13}$ cm$^{-2}$. The data points of the GNP3 and

GNP4 samples with high level of wrinkling ($A_w \sim 50$ %) show also a significant coincidence, however they shift to slightly lower values of $\omega_G$ and $\omega_{G'}$ suggesting moderate overall increase of strain ($\sim -0.02$ %) and net decrease of doping $1.0 \times 10^{13}$ cm$^{-2}$.

Comparing the results of the samples with different level of delamination (schematically shown in Figure 1), the former values of the doping ($1.3 \times 10^{13}$ cm$^{-2}$) suggest that major fraction of the 1-LG is in close contact with the substrate, while the latter show less doping ($1.0 \times 10^{13}$ cm$^{-2}$) due to considerable smaller area of contact of the 1-LG and the SiO$_2$ substrate. The explanation is also coherent with the 1-LG@NPs topography obtained from the AFM data[35].

In analogy to the behavior of the $G_1;G'_1$ pairs, the data points of the $G_2;G'_2$ pairs for the GNP1, GNP2, GNP5 and GNP6 samples with $A_w$ up to $\sim 20$ % coincide in almost a single symmetric spot matching the line of biaxial strain of about 0.12 % with doping within the interval of $0.1 \times 10^{13} - 0.7 \times 10^{13}$ cm$^{-2}$. The data points of the GNP4 sample with high level of wrinkling ($A_w \sim 50$ %) again shift to slightly lower values of $\omega_G$ and $\omega_{G'}$ suggesting moderate increase of strain ($\sim 0.18$ %) with narrower distribution of doping ($\sim 0.2 \times 10^{13}$ cm$^{-2}$).

Plotting finally the $G_2;G'_2$ pairs of the GNP3 sample (obtained by two-component analysis of the G' robust for about 30 % of the spectra) in the correlation diagram, they again coincide with those of GNP4 as for the $G_1;G'_1$ series. It suggests that the general behavior resolving between the 1-LG in contact or delaminated from the SiO$_2$ substrate, is not changed up to the highest wrinkling level. However, due to complex scenario of different strain contributions, especially for high wrinkling regime, the analysis of the G' is not straightforward as that of the G and therefore less suitable for construction of universal dependence of the wrinkling level on parameters obtained from Raman spectra.

As a final point, we subjected the results obtained by analysis of the AFM and Raman data to a general correlation as they represent two independent tools for portrayal of the 1-LG topography.

As suggested by the results of correlation analysis of the Raman spectra, the two components of the G and G' modes can be attributed to different border cases: either to the 1-LG delaminated from the substrate (wrinkles) or 1-LG in contact with the $SiO_2$ substrate. The two regimes are much better resolved in the two subbands of the G mode considering clear difference in doping of the 1-LG. It should be noted that the two dissimilar contributions of the G' also reflect the wrinkling to some extent. In this case, however, the Raman scattering probes two different fractions of the 1-LG; the 1-LG subjected to tension in between the substrate and NPs (corresponding to the $G'_2$) and 1-LG relaxed in wrinkles or moderately compressed ($G'_1$). Thus, the overall balance between the two subbands ($G'_1$ and $G'_2$) also reflects the level of wrinkling.

First, we plotted the relative Raman intensity of the $G_2$ ($I(G_2)/I(G_1) < 1$) against the parameters characterizing NP spatial distribution on the decorated substrates: $N_{NPs}$ and $d_{NP-NP}$. The dependencies are shown in Figure 5, left panel. The $I(G_2)/I(G_1)$ decreases with $d_{NP-NP}$ as $(36.2\pm4.9) \times d^{-1}_{NP-NP} - (0.02\pm0.06)$ and increases linearly with $N_{NPs}$ as $(1.2\pm0.9) \times 10^{-3} N_{NPs} + (0.13\pm0.03)$. The obtained dependencies show rather clear sign of correlation of the delamination due to wrinkling (quantified by the relative intensity of the $G_2$) and spatial distribution of NPs on the substrate.

In order to couple the amount of wrinkles quantified by the AFM to the parameters derived from the Raman spectra, we plotted the relative Raman intensity of the $G_2$ ($I(G_2)/I(G_1) < 1$) against the relative area of wrinkles attributed to the delaminated 1-LG ($A_w$), see Figure 5, right panel. The data fall in a robust dependence, which can be described by a simple linear equation: $I(G_2)/I(G_1) = (1.2\pm0.2) \times 10^{-2} A_w + (0.075\pm0.006)$. We are aware of the fact, that at very low wrinkling regime ($A_w < 5\ \%$), the $G_2$ subband is hardly resolved and the dependence may deviate from the linear one as predicted theoretically[43].

We also applied equivalent procedure to the G' mode and observed a similar linear dependence, which can be expressed as $I(G'_2)/I(G'_1) = (3.4\pm0.6) \times 10^{-3} A_w + 0.103\pm0.015$. The outlier (grey point in the graph) corresponds to the sample with the highest wrinkling level (GNP3) with complex inner structure of the G', which prevented application for the two-component analysis for all spectra in the mapped region. Therefore the universality of the $I(G_2)/I(G_1)$ vs. $A_w$ dependence suggest use of the $G_1$ and $G_2$ integral Raman intensities for reliable quantification of the 1-LG delamination associated with the level of wrinkling in our 1-LG@NPs samples.

In conclusion, we have demonstrated that the amount of wrinkles in 1-LG grown by CVD can be easily controlled by monodisperse NPs, which size is comparable to the strain coherence length. The concentration of the NPs characterized by their average surface density and mean interparticle distance governs the amount of wrinkles created in the 1-LG, characterized by the relative area of the 1-LG corresponding to the delaminated 1-LG in spine of wrinkles, which was determined by advanced processing of the AFM data.

We addressed the doping and strain in 1-LG@NPs by Raman spectra mapping, which enabled analysis of large set of data (900 spectra per sample), sufficiently underlying our conclusions. The Raman spectra exhibit fine structure of the G and G' modes, which can be decomposed in two subbands. Applying the correlation analysis to the Raman shifts of the two components of the principal bands, the obtained data accumulates in two regions of the doping-strain diagram, suggesting presence of two distinct fractions, assigned to the two fractions of the 1-LG. The $G_1$-$G'_1$ pairs, related to the 1-LG interacting with the substrate show significant doping (~$1.3\times10^{13}$ cm$^{-2}$) and a negligible compression (~ 0.05 - 0.1 %), while the $G_2$-$G'_2$ pairs correspond to lower doping (~ $0.2\times10^{13}$ cm$^{-2}$), but larger tensile strain (~ 0.12 – 0.18 %) and they are attributed to the delaminated 1-LG.

The ratio of the integral intensities of the two subbands of both the G and the G' increases linearly with the $A_w$. The general $I(G_2)/I(G_1)$ vs. $A_w$ curve is sufficiently robust within the investigated interval of wrinkling ($A_w$ = 6 – 54 %) and serves as a facile tool for quantification of the level of wrinkling in the 1-LG. In other words, the simple analysis of the integral Raman intensities of the two components of the G mode provides realistic value of the relative delaminated area of the 1-LG.

Our study thus demonstrated that the graphene topography can be both controlled and examined in a rather simple way, which provides a new impulse in strain-engineering of graphene-based nanodevices, site-specific enhancement of the graphene reactivity or investigation of fundamental phenomena related to local gauge fields.

*Acknowledgements:* This work was supported by Czech Ministry of Education, Youth and Sports (ERC-CZ: LL1301) and Czech Science Foundation (15-01953S). We gratefully acknowledge M.P. Morales for providing the monodisperse nanoparticles, prepared within the project MULTIFUN (7RP-262943 and 7E12057). B.P. acknowledges Grant Agency of Charles University (GAUK, no. 1302313).

*Methods*

The preparation of the 1-LG@NPs structures includes synthesis of the 1-LG by CVD and preparation of the NPs coated with oleic acid and dispersed in hexane at proper dilution. The final nanostructure is obtained by transfer of the 1-LG on a SiO$_2$(300nm)/Si substrate decorated with the NPs. The whole procedure is schematically shown in Figure 1.

The graphene samples (left panel in Figure 1) were synthesized using CVD as reported previously. In brief: the Cu foil was heated to 1000 °C and annealed for 20 min under flowing $H_2$ (50 sccm). Then the foil was exposed to $^{12}CH_4$ for 20 min. leaving hydrogen gas on with the same flow rate. The etching of the top layers was realized by switching off the methane and leaving on the hydrogen gas for additional 1-20 min. at 1000°C. Finally the substrate was cooled down quickly under $H_2$.

Monodisperse NPs of iron oxide covered with oleic acid were obtained by decomposition of iron oleate complex in 1-octadecene under inert atmosphere[44]. The real particle size (obtained from the transmission electron microscopy, right panel in Figure 1) was 9.0±0.5 nm (polydispersity index, P.I. = 0.06). Magnetic characterization (see Supplementary, section S2) suggested that the obtained NPs are $\gamma$-$Fe_2O_3$ with very good crystallinity. The NPs were re-dispersed in hexane (Sigma Aldrich, 95 % anhydrous) and their concentration was adjusted to 5.0±0.5 mmol Fe/ml.

The initial dispersion was diluted with pure hexane in order to achieve a representative scale of NP concentration on the substrate. 30 - 50 µl of the adjusted dispersion was spin-coated on the 1cm$^2$ $SiO_2$/Si substrate and spun in the spin-coater for 50 s at 2500 rpm. Then the decorated substrates were dried at subsequent 15 s spanning at 300 rpm. The optimum concentration of the NPs was achieved for the mixture of $10^3 - 10^4$ µl hexane with 1 µl of the initial dispersion (the distribution of the NPs on the substrate was investigated by atomic force microscopy for the dilution range with hexane of $10^1 - 10^5$ µl). The samples with ratios: 1 : $10^3$; 1 : $1.43.10^3$ ; 1 : $2.10^3$; 1 : $3.3.10^3$; and 1 : $10^4$ were used for preparation of the final nanostructures.

The substrates decorated with NP at optimum concentration were annealed for 15 minutes at 300 °C in air to fully carbonize the oleic acid coating (confirmed by XPS). This ensured robust fixing of the NPs on the substrate as tested mechanically by AFM. Finally, the as-grown 1-LG was subsequently transferred on the SiO$_2$/Si substrate with the fixed NPs using polymethylmethacrylate (PMMA), according to reproducible procedures reported previously. Final heat treatment of the samples was carried out at 300 °C in O$_2$ atmosphere.

The AFM images (size of 25 μm$^2$) were captured at ambient conditions in the standard tapping mode with the Veeco Multimode V microscope equipped with the JV scanner, with the resolution of 1024 lines and the scan rate of 0.8 Hz. Fresh RFESP probe ($k$ = 3 N/m, $f_0$ = 75 kHz, nominal tip radius = 8 nm) by Bruker was used for each sample, preserving the wear of the tip comparable.

Alternatively, the samples were studied by the high resolution scanning electron microscope (HR SEM) Tescan Mira 3 LMH with the accelerating potential of 15 kV in 5×10$^4$ - 15×10$^4$ magnification.

Both the AFM and HR SEM images were processed in order to reveal the $d_{NP}$, distance $d_{NP-NP}$ and $N_{NPs}$, which were compared to the expected values determined by a simple simulation of random distribution of the NPs on flat substrate carried out in the MATLAB network.

The AFM images were further subjected to advanced analysis with the help of Gwyddion software[45]. The values of $\sigma_p$[46, 47] of the bare substrates, decorated substrates, smooth and delaminated parts of the 1-LG were determined. The value of the relative area attributed to delaminated 1-LG, $A_w$ was finally extracted for each image. Complete description of the procedure can be found in the recent work of Pacakova et al[35].

The Raman spectra were acquired by a LabRam HR spectrometer (Horiba Jobin-Yvon) using 633 nm He/Ne excitation. The spectral resolution was about 1 cm$^{-1}$. The spectrometer was interfaced to a microscope (Olympus, 100 x objectives) so the spot size was about 1μm$^2$. The typical laser power measured at the sample was about 1 mW. The Raman maps (size of 30 x 30 μm$^2$) were recorded with step of 1 μm. All bands were fitted either as a single pseudo-Voigt functions, $\Omega(I,\omega)$ or their superposition; for explicit form of the relation, see part S.4 in Supplementary. The Raman maps (containing typically 900 spectra) were analysed using custom made routine in Octave enabling automated fitting in batches and setting of relevant constrains between the parameters.

*Figures*

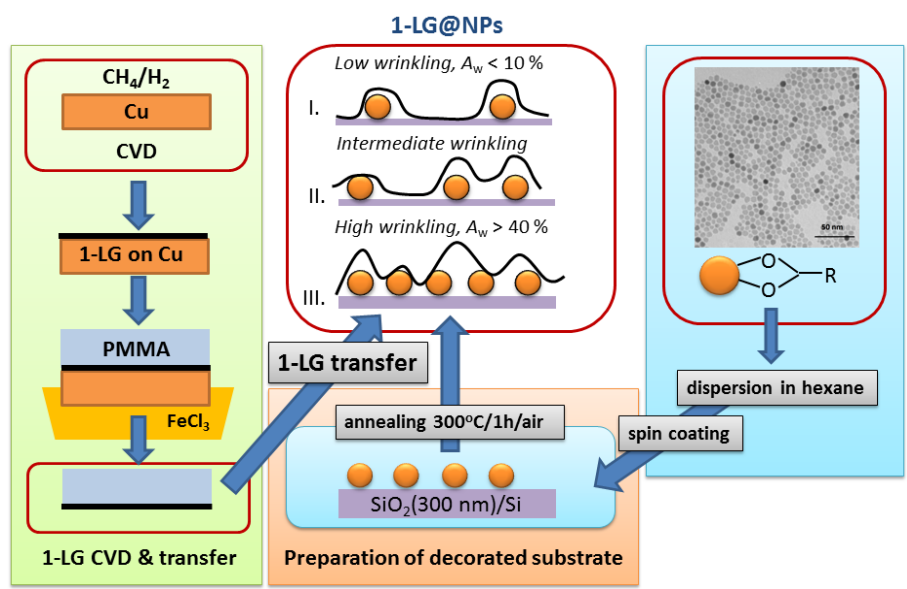

Figure 1

Preparation scheme of the 1-LG@NPs samples. The single-layer graphene (1-LG) was obtained by CVD technique (left panel, green) and transferred with the help of PMMA on $SiO_2$(300nm)/Si substrate decorated with fixed monodisperse (9±0.5 nm) iron oxide nanoparticles (NPs). The density of NPs was adjusted by initial concentration of the dispersion of NPs in hexane, which was deposited on the substrate by spin-coating (bottom panel, orange). The NPs were fixed on the substrate by carbonization of the organic coating (oleic acid). The oleic acid coated NPs were obtained by decomposition of organic precursor in high boiling solvent (right panel, blue). The final 1-LG@NPs nanostructures, obtained after removal of the PMMA are shown in the middle top panel and form three limit cases: I. Low wrinkling regime - 1-LG sheet is in contact with the substrate in the area between the NPs (low NP density), II. Intermediate wrinkling regime – less of the 1-LG is in contact with the substrate, the layer is partially delaminated and III. High wrinkling regime - 1-LG sheet is significantly delaminated from the substrate (high NP density).

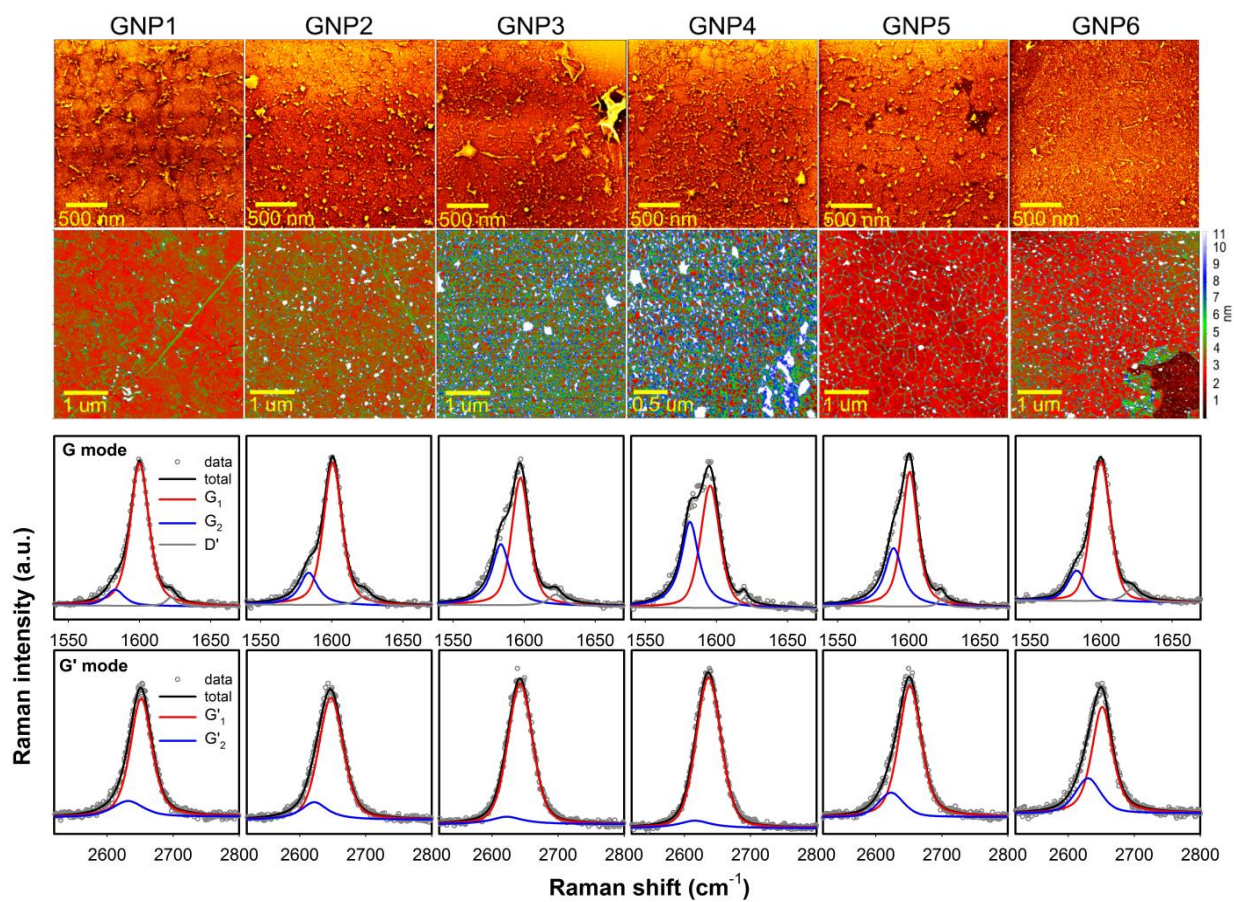

Figure 2

Typical AFM (first row) and HR SEM (second row) images of the 1-LG@NPs samples (GNP1 – GNP6). The morphology of the 1-LG layer shows characteristic wrinkles, which density increases with increasing concentration of the NPs decorating the $SiO_2$/Si substrate. Typical Raman spectra of the 1-LG@NPs samples (GNP1 – GNP6) in the G and G' region are presented in the last two rows. The experimental data are marked with dark grey open circles, the fit of the individual components of the G and G' modes is shown by red, blue and dark grey for the $G_1$, $G_2$ and D' and by red and blue for the $G'_1$ and $G'_2$, respectively. The resulting curve (sum of the individual components) is represented by solid black line.

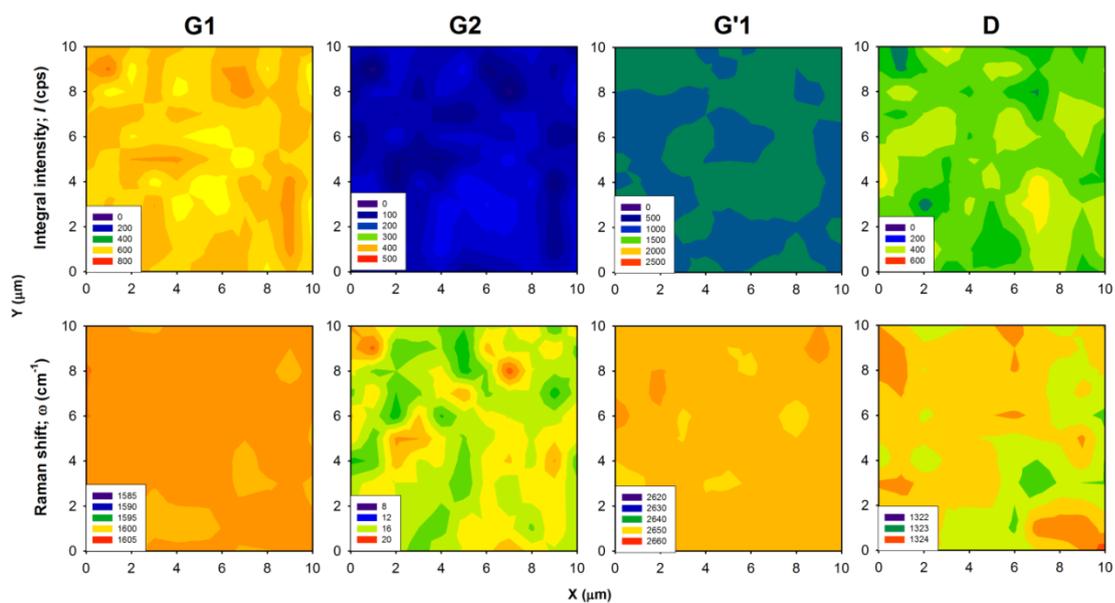

Figure 3

Typical Raman maps (size of 10 x 10 μm$^2$) of selected principal Raman active modes for the GNP3 sample; the same maps of the other samples are included in Supplementary (Figures S.1.2 – S.1.6.). The upper row corresponds to the integral intensity and the bottom row to the Raman shift, ω of the $G_1$, $G_2$, $G'_1$ and D modes, respectively.

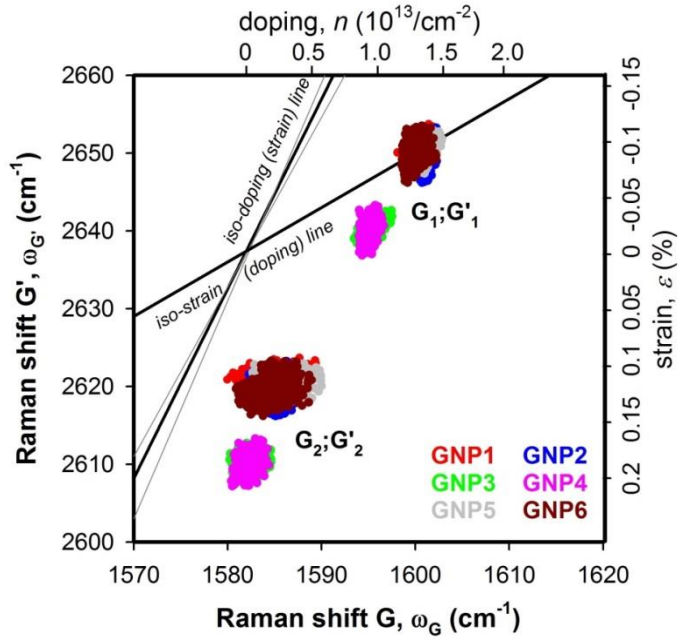

Figure 4

Correlation plot of the Raman shift, ω of the principal Raman modes in 1-LG@NPs samples. The $\omega_{G1}$, $\omega_{G'1}$ and $\omega_{G2}$, $\omega_{G'2}$ correlations are distributed in different areas of the plot suggesting different level of strain and doping with respect to the character of the 1-LG, which is either in contact with the substrate and shows higher doping and lower strain level ($G_1$, $G'_1$) or forms a wrinkle ($G_2$, $G'_2$) resulting in lower doping but larger tension. The dashed grey lines with intersection at $P_0$ (1582, 2637) cm$^{-1}$ correspond to the border lines for zero doping and biaxial strain with the slope of 2.45 and 0.7, respectively. The thin grey lines represent a spread of the iso-doping line of the biaxial strain with the border values of the slope found in literature (2.25 - 2.8). The absolute values of the doping, $n$(cm$^{-2}$) and the strain, $\varepsilon$(%) are depicted on the top and on the right edge of the graph, respectively.

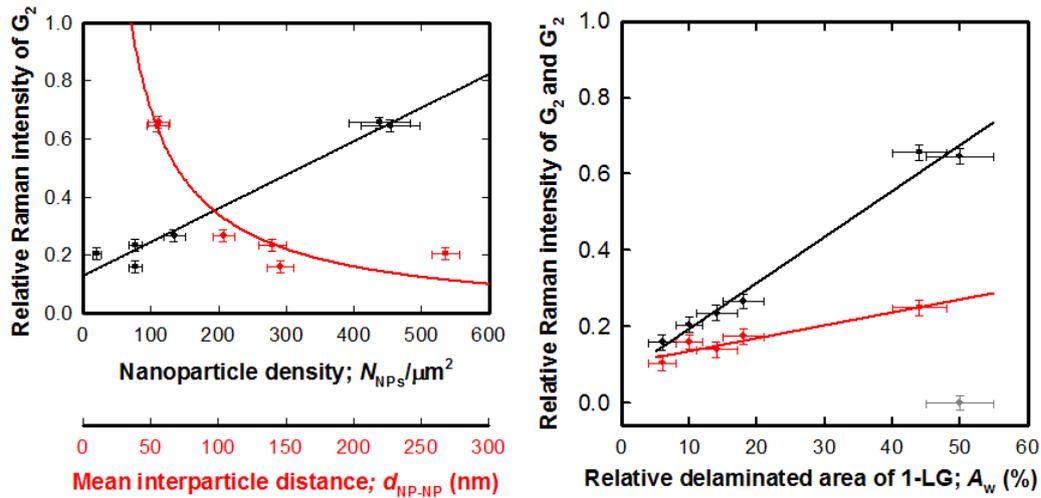

Figure 5

Correlation of the parameters representing the spatial distribution of nanoparticles and level of wrinkling of the 1-LG layer. The left panel shows the mean nanoparticle density, $N_{NPs}$ (black points) and mean interparticle distance, $d_{NP-NP}$ (red points) in context of the relative delaminated area, $A_w$ determined by the analysis of the AFM data. The data roughly scales with $N_{NPs}$ and $1/d_{NP-NP}$ for increasing relative intensity of the $G_2$. The right panel correlates the $A_w$ (obtained from AFM) to the ratio of the integral intensity of the $G_2$ (attributed to the delaminated fraction of the 1-LG) and $G_1$ modes (black points), which results in a robust linear dependence. Analogues plot is shown for the pair of $G'_1$ and $G'_2$ modes (red points), which follows the same trend with much lower slope. The grey outlying point corresponds to the sample with the largest $A_w$ values (GNP3), which G' is better analyzed by using a single peak with larger FWHM.

*Cover Figure*

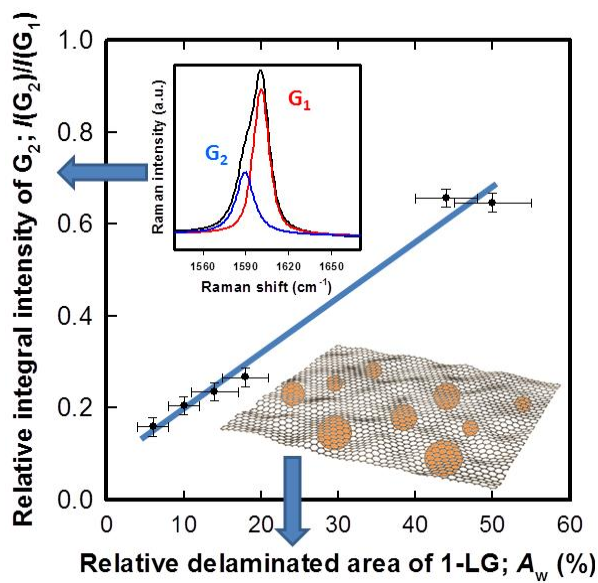

# Supplementary

**Graphene wrinkling induced by monodisperse nanoparticles: facile control and quantification**

Jana Vejpravova[1]*, Barbara Pacakova[1], Jan Endres[2], Alice Mantlikova[1], Tim Verhagen[1], Vaclav Vales[3], Otakar Frank[3] and Martin Kalbac[3]**

[1]Institute of Physics AS CR, v.v.i., Department of Magnetic Nanosystems, Na Slovance 2, 18221 Prague 2, Czech Republic

[2]Charles Univeristy in Prague, Faculty of Mathematics and Physics, Department of Condensed Matter Physics, Ke Karlovu 5, 12116 Prague 2, Czech Republic

[3]JH Institute of Physical Chemistry AS CR,v.v.i., Dolejskova 3, 18200 Prague 8, Czech Republic

*vejpravo@fzu.cz, **martin.kalbac@jh-inst.cas.cz

## S.1. Additional results of Raman mapping

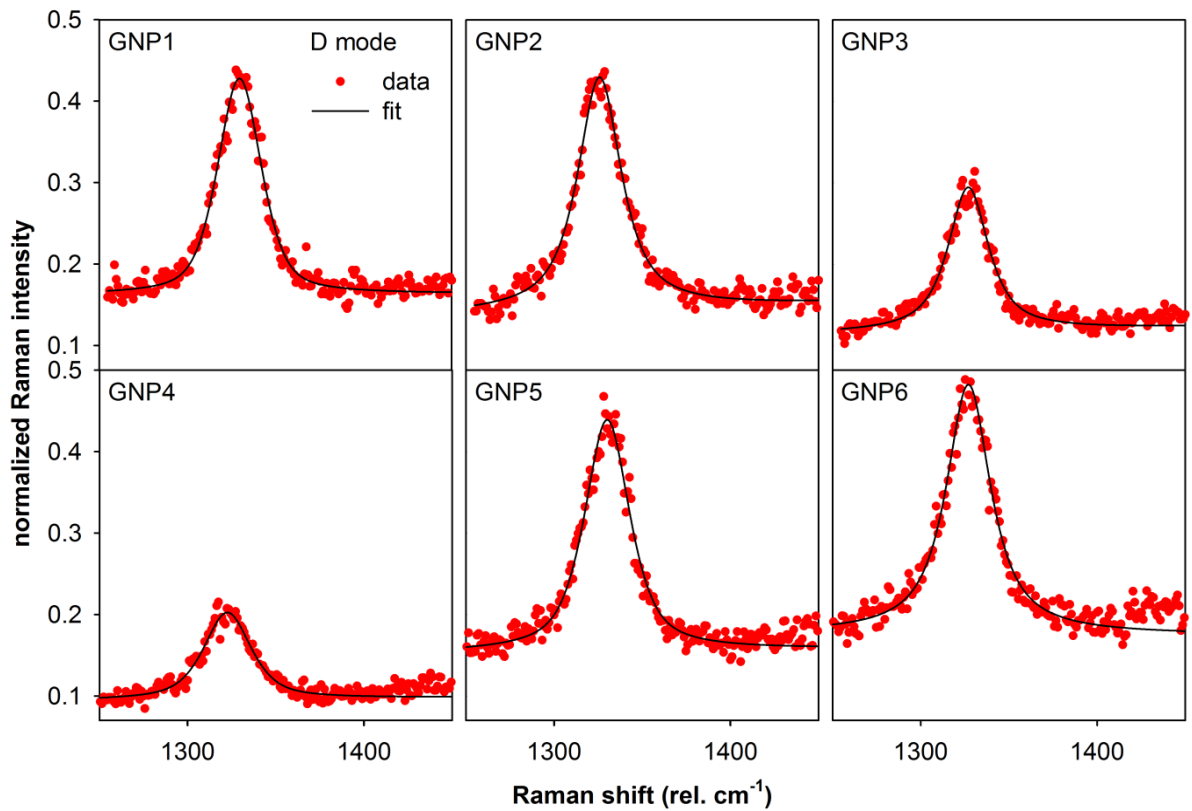

S1.1. Typical Raman spectra of the GNP1-6 samples in the D-mode region together of the fit by a single pseudo-Voigt function.

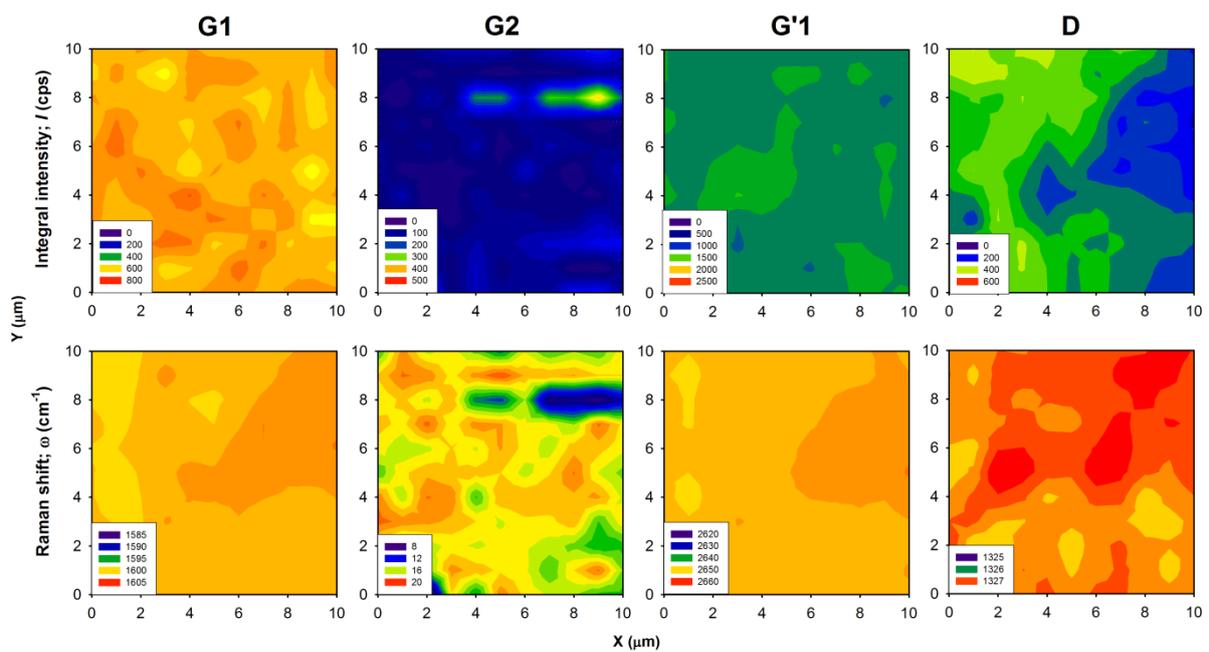

S1.2. Raman maps of Raman shift and integral intensity of the principal graphene modes for the GNP1 sample.

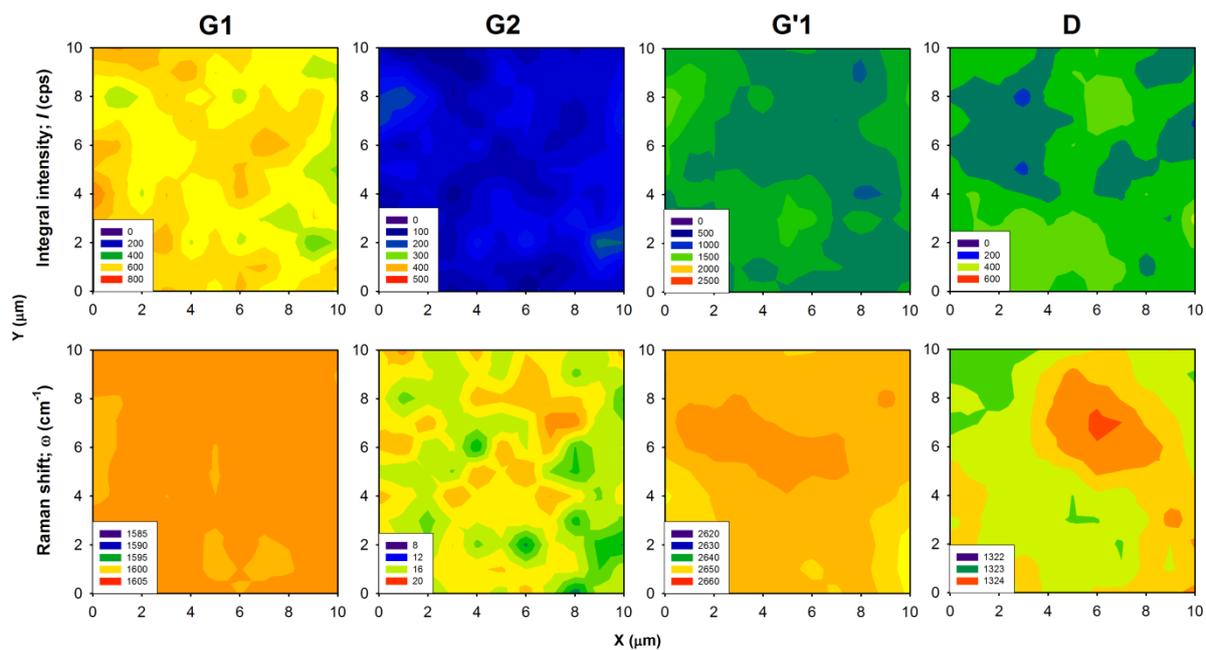

S1.3. Raman maps of Raman shift and integral intensity of the principal graphene modes for the GNP2 sample.

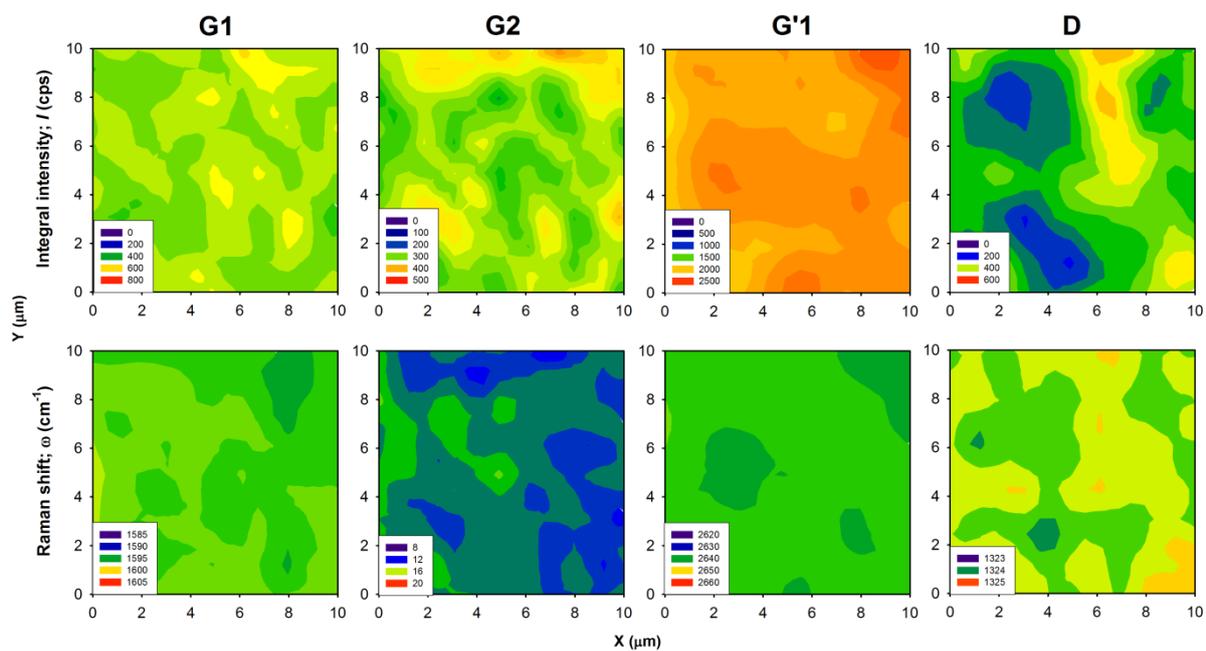

S1.4. Raman maps of Raman shift and integral intensity of the principal graphene modes for the GNP3 sample.

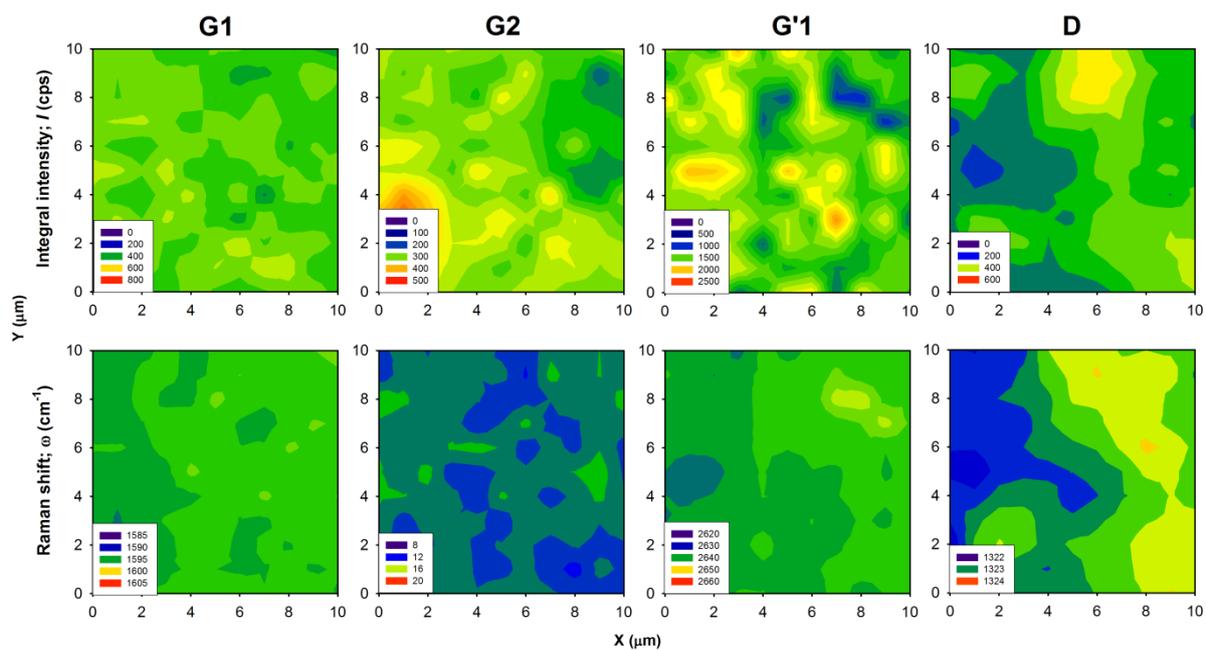

S1.5. Raman maps of Raman shift and integral intensity of the principal graphene modes for the GNP4 sample.

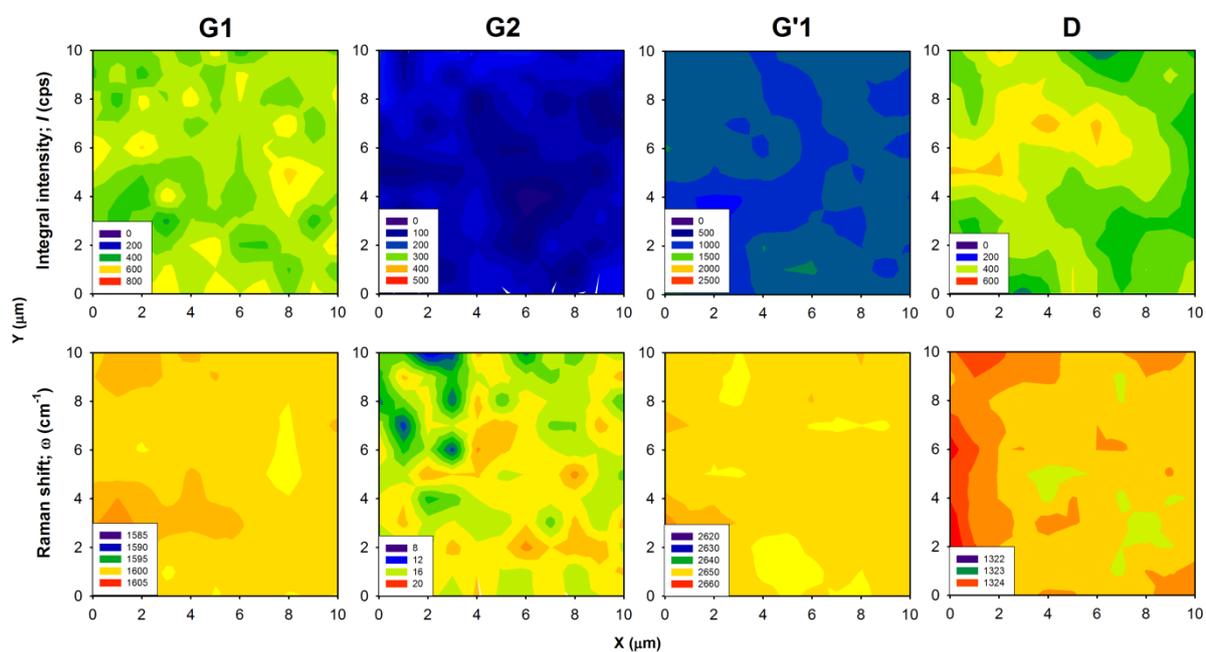

S1.6. Raman maps of Raman shift and integral intensity of the principal graphene modes for the GNP6 sample.

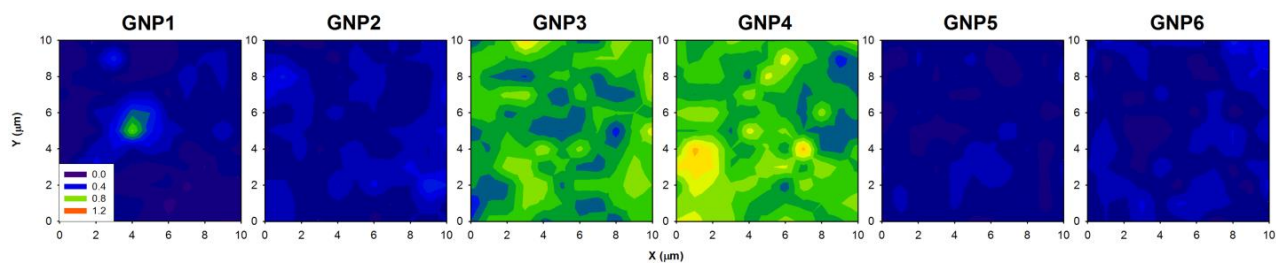

S1.7. Spatial distribution of the relative intensity of the $G_2$ with respect to the $G_1$ mode for GNP1 – GNP6 samples.

Table S1.1. Basic parameters obtained from analysis of the fine structure of the G and G' mode: FWHM – full width at half maxima and $\alpha$ - fraction of the Lorentzian component. For the GNP3, the best match of the G' was achieved for a single pseudo-Voigt peak with larger FWHM, therefore the $\alpha$ G'$_2$ is just estimation from the less significant fit.

| Sample | FWHM G$_1$ (cm$^{-1}$) | FWHM G$_2$ (cm$^{-1}$) | $\alpha$ G$_1$ | FWHM G'$_1$ (cm$^{-1}$) | $\alpha$ G'$_1$ | $\alpha$ G'$_2$ |
|---|---|---|---|---|---|---|
| **GNP1** | 14.8±0.8 | 15.0±0.2 | 0.69±0.11 | 39.5±1.5 | 0.60±0.04 | 0.57±0.32 |
| **GNP2** | 14.8±0.6 | 15.0±0.2 | 0.53±0.10 | 43.0±1.8 | 0.51±0.07 | 0.77±0.23 |
| **GNP3** | 15.2±0.4 | 15.0±0.2 | 0.68±0.22 | 45.4±1.0 | 0.42±0.05 | 0.50±0.20* |
| **GNP4** | 15.9±0.4 | 15.0±0.2 | 0.43±0.16 | 44.0±1.4 | 0.35±0.05 | 0.98±0.02 |
| **GNP5** | 13.6±0.4 | 15.0±0.2 | 0.65±0.22 | 42.6±1.0 | 0.53±0.06 | 0.78±0.20 |
| **GNP6** | 15.2±0.4 | 15.0±0.2 | 0.68±0.22 | 42.4±1.2 | 0.49±0.04 | 0.66±0.18 |

## S.2. Magnetic characterization of the nanoparticles

We performed basic characterization of magnetic properties of the dried NP sample. The temperature dependence of the zero-field-cooled and field-cooled magnetization show characteristic saturation of the FC curve, typical for strongly interacting system of superparamagnetic NPs. Further, the refinement of un-hysteretic loops was carried out in order to determine the median magnetic moment and the so-called magnetic size of the NPs. The mean size fraction corresponds to app. 10 nm large NPs, which agrees well with the values obtained from AFM and SEM and hence suggests excellent crystallinity of the NPs.

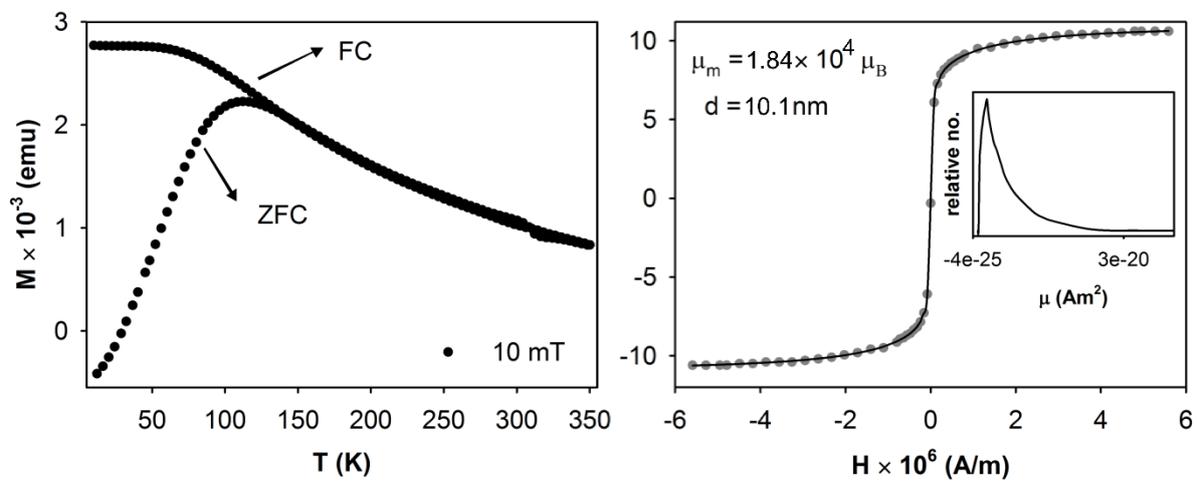

S2.1. Temperature dependencies of ZFC and FC magnetization of the NP sample, together with refinement of the un-hysteretic curves in the SPM state. Distribution of magnetic moments is shown in the inset. Values of the mean magnetic moments and magnetic diameter are also depicted in the image

## S.3. Additional characterization of the nanoparticles and GNP1-GNP6 samples by HR SEM and AFM

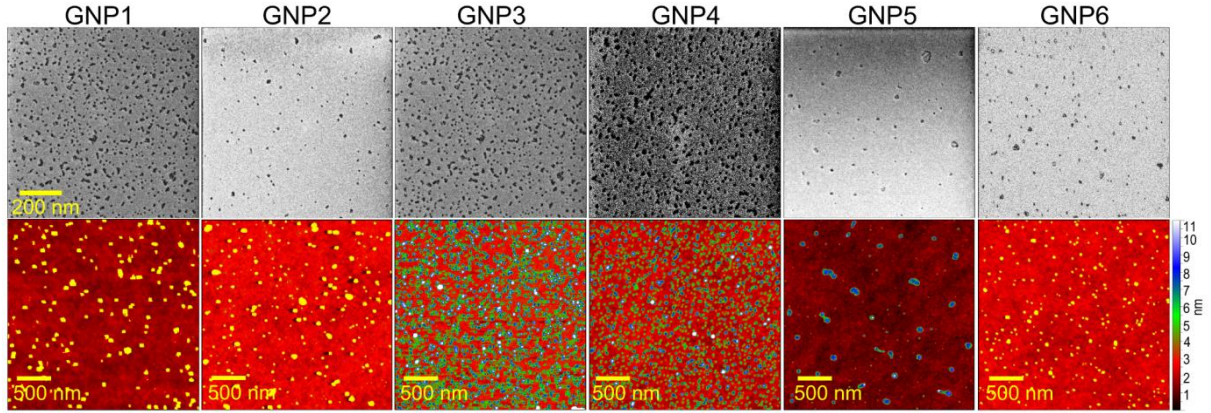

S3.1. High-resolution SEM images of the nanoparticles dispersed on Si/SiO$_2$ substrate (top) and example of AFM images of the substrate Si/SiO$_2$ decorated with NPs (bottom).

Table S3.1. Basic parameters obtained from analysis of the HR SEM and AFM imaging: nanoparticle density, $N_{NPs}$; mean interparticle distance, $d_{NP-NP}$; wrinkled area of the 1-LG, $A_w$;

| Sample | $N_{NPs}/\mu m^2$ | $d_{NP-NP}$ (nm) | $A_w$ (%) |
|---|---|---|---|
| **GNP1** | 77±5 | 146±5 | 6±2 |
| **GNP2** | 77±5 | 140±5 | 14±3 |
| **GNP3** | 454±45 | 54±4 | 50±5 |
| **GNP4** | 438±44 | 54±4 | 44±5 |
| **GNP5** | 20±3 | 268±2 | 10±3 |
| **GNP6** | 134±8 | 104±4 | 18±3 |

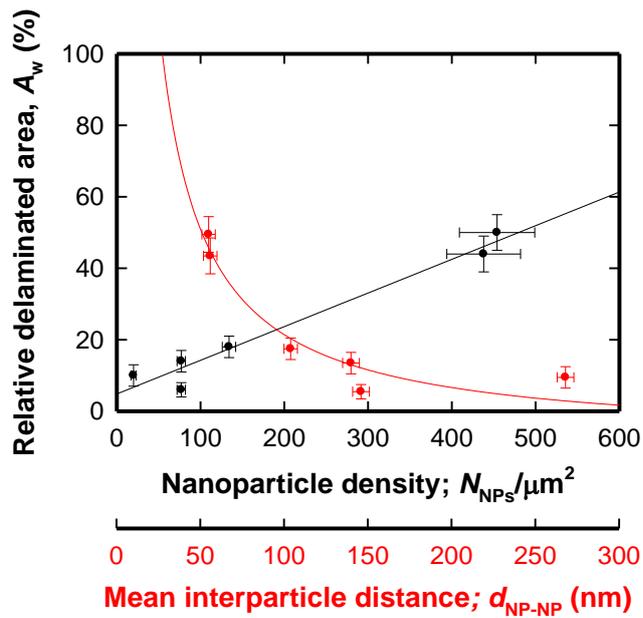

S3.2. Correlation of the mean nanoparticle density, $N_{NPs}$ (black) and mean interparticle distance, $d_{NP-NP}$ (red) to relative delaminated area of 1-LG, $A_w$ determined from AFM data. The $A_w(N_{NPs})$ dependence can be expressed as a linear function: $A_w(N_{NPs}) = a(N_{NPs}) + b$, where a = 0.094±0.009 and b = 4.8±2.4; the $A_w(d_{NP-NP})$ dependence follows approximately a hyperbolic function: $A_w(d_{NP-NP}) = c/(d_{NP-NP}) + d$, where c = 2964±389 and d = 7.6±4.6.

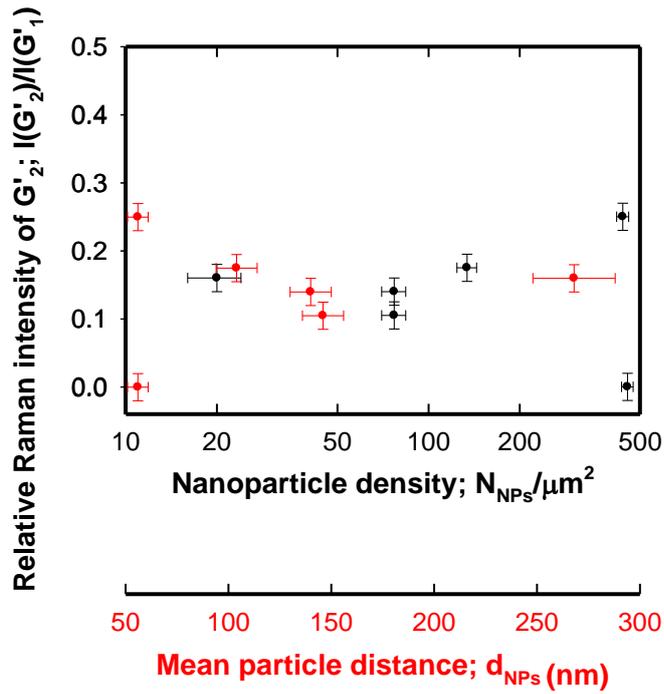

S3.3. Correlation of the key parameters representing the spatial distribution of nanoparticles and level of wrinkling of the 1-LG layer estimated from the analysis of the G' mode (mean nanoparticle density, $N_{NPs}$ (and mean interparticle distance, $d_{NP-NP}$ (red), relative area of wrinkles, $A_w$). The dependencies do not show a clear monotonic trend as in case of the G mode-related features due to complex structure of the G' mode.

## S.4 Profile analysis of the Raman spectra

The individual Raman peaks were fitted by the profile function $\Omega(I,\omega)$ (eq.S1) aproximated in the form of the pseudo-Voigt function (linear combination of the Gaussian and Lorentizan as a sufficient approximation of their convolution – Voigt function). The symbols used in equation S4.1 have the following meaning: $I$ - Raman intensity, $\omega$ - Raman shift, $\omega_0$ - peak position, $\Gamma$ - full width at half maximum of the peak and $\alpha$ - fraction of the Lorentzian component. The Gaussian component serves as a measure of distribution of the peak parameters due to finite size of the laser spot (~1 μm$^2$), which is expected to be about one order larger then the local variation of the parameters at nm scale.

$$\Omega(I,\omega) = (1-\alpha)I\sqrt{\frac{\ln 2}{4\pi\Gamma^2}}\exp\left[\frac{-\ln 2(\omega-\omega_0)^2}{4\Gamma^2}\right] + \alpha\left[\frac{1}{2\pi\Gamma}\frac{I}{1+\frac{(\omega-\omega_0)^2}{4\Gamma^2}}\right] \qquad (S4.1)$$